\documentclass[aps,twocolumn,amsmath,amssymb,showpacs,showkeys,floatfix,a4paper,nofootinbib]{revtex4}

\usepackage{epsfig}
\usepackage{amssymb}
\usepackage[utf8]{inputenc}
\usepackage{soul,color}
\usepackage{graphicx}
\usepackage{dcolumn}
\usepackage{bm}

\usepackage{float}
\usepackage{mathrsfs}
\usepackage{tabularx}
\usepackage{midfloat}
\usepackage[normalem]{ulem}
\newcommand{\be}{\begin{equation}}
\newcommand{\ee}{\end{equation}}
\newcommand{\bea}{\begin{eqnarray}}
\newcommand{\eea}{\end{eqnarray}}
\newcommand{\ba}{\begin{array}}
\newcommand{\ea}{\end{array}}
\newcommand{\bi}{\begin{itemize}}
\newcommand{\ei}{\end{itemize}}
\newcommand{\lan}{\langle}
\newcommand{\ran}{\rangle}

\begin{document}

\title{Heavy Ion Double Charge Exchange  Reactions as Probes for Two-Body Transition Densities}

\author{Jessica I.Bellone$^{1,2}$ {\thanks{Electronic address: jessica.bellone@ct.infn.it}}}
\author{Maria Colonna$^{1}$ {\thanks{Electronic address: colonna@lns.infn.it}}}
\author{Danilo Gambacurta$^{1}$ {\thanks{Electronic address: gambacurta@lns.infn.it}}}
\author{Horst Lenske$^{3}${\thanks{Electronic address: horst.lenske@theo.physik.uni-giessen.de}}}

\affiliation{
$^1$INFN-LNS, I-95123 Catania, Italy
\\
$^2$Universit\'a degli studi di Catania, Dipartimento di Fisica e Astronomia ``E. Majorana'', via Santa Sofia, 64, 95123 Catania, Italy\\
$^3$Institut f\"{u}r Theoretische Physik, Justus-Liebig-Universit\"{a}t Giessen, D-35392 Giessen, Germany 
}

\date{\today}

\begin{abstract}
\begin{center}
(NUMEN Collaboration)
\end{center}
Collisional heavy ion double charge exchange (DCE) reactions, induced by second order nucleon-nucleon interactions, are shown to provide access to the two-body transition densities  of the complementary DCE transitions in the interacting nuclei. 
Corresponding two-body operators are introduced, treating the second order distorted wave reaction amplitude in the s-channel interaction form.
The theoretical results are applied to the reaction $^{18}O+{}^{76}Se\to {} ^{18}Ne+{}^{76}Ge$ at $T_{lab}=270$~MeV, being $^{76}Ge$ a candidate for neutrino--less double beta decay.
\end{abstract}
\maketitle

{\it Introduction.}
Second-order processes provide a crucial touchstone to test our understanding
of the underlying physics mechanisms and
of our hierarchically-conceived conceptual modelling of physics phenomena.
Typical examples are the double-gamma decay in quantum-electrodynamics \cite{Kramp1987,Waltz2015,Soderstrom2020} and the double-beta decay (DBD), with or without neutrinos, in the electro-weak case \cite{Engel,Avignone,Ejiri,Tomoda:1990rs,
Martinez2013,
Kotila:2021mtq,Nitescu:2024rsh},
which are probes of non-hadronic origin. In early investigations with hadron beams, pion-induced Double Charge-Exchange (DCE) reactions were used for spectroscopic investigations  of second order spin-isospin nuclear excitations, see e.g.  
\cite{Siciliano:1986kj,Siciliano:1990yf,
Johnson:1993lsj,Gilman:1986rg}
and also 
\cite{Auerbach:1987gn,Auerbach:1988ir,Auerbach:1990vh,Auerbach:1996tb}, but the results were elusive for various reasons \cite{Haxton:1997rn}. 
Also, first studies with nuclear beams did not lead to conclusive results
\cite{blo1995plb362,dra1980prl45,nau1982prc25,das1986prc34,bes1983npa405}. A new opportunity for extending the investigations into the sector of nuclear interactions has become available through recent research activities on
heavy ion DCE reactions~\cite{Ejiri:2022ujl,Cappuzzello:2022ton}, 
inducing in a complementary manner $A(N,Z)+a(N',Z') \to B(N\pm 2,Z\mp 2)+b(N'\mp 2,Z'\pm 2)$  transitions in the interacting nuclei.

Heavy ion DCE reactions proceed in general by two vastly different reaction mechanisms, namely soft sequential transfer processes,
reflecting single nucleon and nucleon pair mean-field spectroscopy, and hard collisional charge exchange processes
by nucleon-nucleon (NN) interactions. The latter produce daughter nuclei which spectroscopically correspond to two particle--two hole configurations relative to the parent nuclei,
probing directly nuclear isovector spectroscopy by operators as encountered in two-neutrino
and neutrinoless double beta decay \cite{Lenske:2021jnr,Cappuzzello:2022ton,Lenske:2024dsc}.
A central issue of this Letter is to point out that collisional DCE reactions are a highly valuable tool
to study explicitly nuclear spectroscopy induced by short--range two--body nuclear interactions, thus opening
a new window to investigate the rarely - if ever - studied response functions of higher order many--body operators.
More specifically, we focus on the {still little known} spin-isospin two--body transition densities (2BTD). For that kind of research collisional DCE reactions mediated by isovector NN interactions are of advantage by two reasons: First, the final states are defined unambiguously by their charge and mass as a clear signature for a two--body nuclear structure event. Second, the underlying interactions are well known from NN scattering and from light (and heavy) ion Single Charge Exchange (SCE) reactions \cite{Lenske:2019cex}.
Nuclear DCE transitions can be depicted as driven by effective rank-2 isotensor two--body interactions created dynamically during the short period in which the nuclei are in close contact. These effective DCE interactions are extremely elusive, existing on the time scale of direct nuclear reactions, $\tau_{DR}\sim 10^{-22}$~sec. Since short pulses of this kind
lead to broad energy distributions of spectroscopic strength, heavy ion DCE reactions are the perfect tool to explore the poorly known landscape of 2BTD.

\begin{figure}
\begin{center}
\includegraphics[width = 5cm]{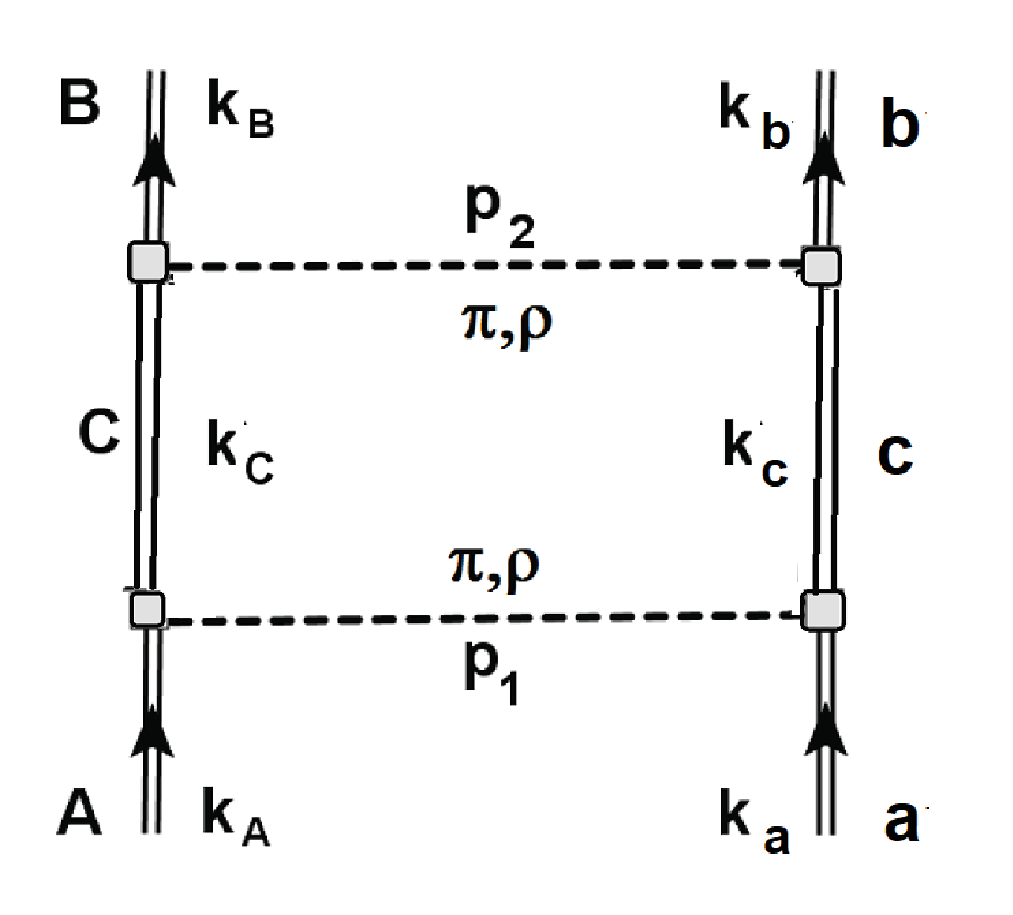}
\caption{The DSCE ladder diagram. The four--momenta of the incoming ($A,a$) and outgoing ($B,b$) nuclei are denoted by $k_{A,a}$ and $k_{B,b}$, respectively; those of the intermediate SCE--excited nuclei $C,c$  are $k_{C,c}$. The DSCE process, mediated by the isovector NN T-matrix, is indicated by the sequential exchange of $\pi$- and $\rho$-mesons.
See text for further discussions.}
\label{fig:DSCE}
\end{center}
\end{figure}

{\it Aspects of DSCE Reaction Theory.}
The DCE reactions of interest are peripheral, grazing reactions. Nuclear direct reaction (DR) theory provides the tools to handle theoretically a reaction
occurring on top of a background of a multitude of other reaction channels.
The essence of DR
theory is a separability \emph{ansatz} by expanding the total wave function $\Psi^{(+)}_{aA}$  in terms of configurations
$\chi^{(\pm)}_\gamma|a_\gamma A_\gamma\ran$ of asymptotic nuclear eigenstates $|a_\gamma\ran\in \{a\}$ and $|A_\gamma\ran\in\{A\}$.
The expansion coefficients are given by the distorted waves $\chi^{(\pm)}_\lambda(\mathbf{r}_\lambda,\mathbf{k}_\lambda)$
of relative motion, depending on the distance $\mathbf{r}_\lambda$ and the invariant relative three-momentum $\mathbf{k}_\lambda$
in partition $\lambda$. As depicted in Fig.\ref{fig:DSCE}, the reaction proceeds from the incident channel $\alpha= a+A$ by
passing through a set of intermediate SCE channels $\gamma = c+C$ to the DCE exit channel $\beta=b+B$. The invariant
energy $\omega_\alpha=\sqrt{s_{aA}}$, $s_{aA}=(k_a+k_A)^2$, available for the reaction is defined by the four-momenta
$k_{a,A}$ of the incoming ions. In the ion--ion rest frame we find $k_{a,A}=(E_{a,A},\pm \mathbf{k}_\alpha)$. 

The double sequential CE (DSCE) reaction proceeds by acting twice with the isovector (T=1) NN T-matrix $\mathcal{T}_{NN}$, indicated in Fig.\ref{fig:DSCE}, by virtual $\pi$- and $\rho$-meson exchange. In the rest frame the exchanged momenta $p_{1,2}=(0,\mathbf{p}_{1,2})$
are space-like. The DSCE reaction amplitude is of second order in $\mathcal{T}_{NN}$. Taking into account 
{initial state (ISI) and final state (FSI) interactions,} a formulation in
second-order distorted wave (DW) theory is the proper approach:
\bea\label{eq:MDSCE}
&&M^{(2)}_{\alpha\beta}(\mathbf{k}_\alpha,\mathbf{k}_\beta)=\sum_{\gamma=c,C}\int \frac{d^3k_\gamma}{(2\pi)^3}\\
&&\lan \chi^{(-)}_\beta|\mathcal{F}_{\beta\gamma}|\chi^{(+)}_\gamma\ran
G^{(+)}_\gamma(\omega_\alpha)\lan\widetilde{\chi}^{(+)}_\gamma|\mathcal{F}_{\gamma\alpha}|\chi^{(+)}_\alpha\ran ,\nonumber
\eea
where $\widetilde{\chi}^{(+)}_\gamma$ is a dual wave vector.
The SCE sub-processes are described by two--body transition form factors, e.g.
$\mathcal{F}_{\gamma\alpha}(\mathbf{r}_\alpha)=\lan cC|\mathcal{T}_{NN}|aA\ran$ and $\mathcal{F}_{\beta\gamma}(\mathbf{r}_\beta)$
is defined accordingly.

In non-relativistic reduction, the
intermediate channel propagator is
$G^{(+)}_\gamma\approx (T_\alpha+\Delta M_{\alpha\gamma}-T_\gamma-\varepsilon_\gamma+i\eta)^{-1}$ where $T_{\alpha,\gamma}$ are
the kinetic energies, $\Delta M_{\alpha\gamma}=M_a+M_A-M_c-M_C$ denotes the ground state mass difference, and $\varepsilon_\gamma$ is
the sum of excitation energies. The propagator is strongly dominated by the pole at
$E_\gamma=T_\gamma+\varepsilon_\gamma=T_\alpha + \Delta M_{\alpha\gamma}$. The complete propagator, including the
intermediate distorted waves, approaches a delta distribution $\sim \delta^3(\mathbf{r}_\alpha-\mathbf{r}_\beta)$,
see \cite{Lenske:2024dsc}.  Anticipating that result (see  Section I of the Supplemental Material \cite{SM} for more details),
we average $G^{(+)}_\gamma$  over an interval $\Delta E_\gamma$ around the pole at $E_\gamma$. The resulting
(logarithmic function) $L^{(+)}_\gamma(\omega_\alpha)$ is independent of $\mathbf{k}_\gamma$.
In this \emph{mean-energy approximation} and using the momentum representation
\cite{Lenske:2018jav,Lenske:2021jnr,Lenske:2024dsc}
the DSCE amplitude attains the appealing form
\bea
M^{(2)}_{\alpha\beta}(\mathbf{k}_\alpha,\mathbf{k}_\beta)\approx
\int d^3p_1\int d^3p_2 
D_{\alpha\beta}(\mathbf{p}_1+\mathbf{p}_2)\Pi_{\alpha\beta}(\mathbf{p}_1,\mathbf{p}_2).\nonumber
\eea
ISI and FSI are contained now in the distortion amplitude $D_{\alpha\beta}(\mathbf{q})=\frac{1}{(2\pi)^3}\lan \chi^{(-)}_\beta|e^{i\mathbf{q}\cdot \mathbf{r}_\alpha}|\chi^{(+)}_\alpha\ran$, thus appearing as renormalization factors with the effect of quenching the polarization tensor
\be
\Pi_{\alpha\beta}(\mathbf{p}_1,\mathbf{p}_2)=
\sum_{\gamma=c,C}L^{(+)}_\gamma(\omega_\alpha)\mathcal{F}_{\beta\gamma}(\mathbf{p}_2)\mathcal{F}_{\gamma\alpha}(\mathbf{p}_1)
\ee
which collects the bare combined spectroscopic strength in the two nuclei before ISI/FSI renormalization.

\textit{The Polarization Tensor.}
The building blocks of $\mathcal{T}_{NN}$ and transition form factors 
are the one--body isovector operators
$\mathcal{R}^{\pm}_S(\mathbf{p})=e^{i\mathbf{p}\cdot \mathbf{r}} \bm{\sigma}^S\tau_\pm$ of spin-scalar (S=0) and spin-vector (S=1) character  \cite{Lenske:2018jav}.
$\mathcal{T}_{NN}$ is given by a superposition of scalar products of these operators, one acting in the $\{a\}$ and the other in the $\{A\}$ system.
Considering for simplicity at this formal stage only rank-0 central isovector (T=1) interactions with vertex functionals $V_{ST}(\mathbf{p})$ (see Supplemental Material \cite{SM}, Section II), we find
\bea
\Pi_{\alpha\beta}(\mathbf{p}_1,\mathbf{p}_2)=
\sum_{S_1,S_2}V_{S_2T}(\mathbf{p}_2)~\Pi^{(S_2S_1)}_{\alpha\beta}(\mathbf{p}_2,\mathbf{p}_1)~V_{S_1T}(\mathbf{p}_1) \nonumber,
\eea
with the spin--spin tensors
\bea\label{eq:NucTensor}
&&\Pi^{(S_2S_1)}_{\alpha\beta}(\mathbf{p}_1,\mathbf{p}_2)=\sum_{\gamma=cC} L^{(+)}_\gamma(\omega_\alpha)
\\
&&{F^{(BC)}_{S_2}(\mathbf{p}_2)\cdot F^{(bc)}_{S_2}(\mathbf{p}_2)F^{(ca)}_{S_1}(\mathbf{p}_1)\cdot
F^{(CA)}_{S_1}(\mathbf{p}_1)}\nonumber.
\eea
Contractions over spin components are indicated by scalar products. The
one--body transition form factors 
are matrix elements of the isovector operators $\mathcal{R}^\pm_S$, e.g. $F^{CA}(\mathbf{p})=\lan C|\mathcal{R}^\pm_S(\mathbf{p})|A\ran$. They are one-body transition densities (1BTD)  in momentum representation.

{\it DCE form factors and 2BTD in s--channel formalism.} In the standard t-channel representation, depicted in Eqs.\eqref{eq:MDSCE}-
\eqref{eq:NucTensor},
the DSCE reaction amplitude is obtained by coupling projectile and target SCE transitions separately in each interaction step \cite{Bellone:2020lal}. Hence, the t-channel formalism is not suitable to extract directly the desired spectroscopic information on nuclear matrix elements connecting the initial state and the final DCE states \cite{Lenske:2021jnr,Lenske:2024dsc}.
The s--channel presentation is obtained by a rotation in (spin) angular momentum space by which the nuclear polarization tensor, Eq.\eqref{eq:NucTensor}, becomes
\begin{equation}
 \Pi^{(S_2S_1)}_{\alpha\beta}(\mathbf{p}_1,\mathbf{p}_2) =
\sum_{\gamma=cC}L^{(+)}_\gamma(\omega_\alpha)
\sum_{S,M}(-)^{S_1+S_2-S+M}
\label{eq_PT}
\end{equation}
\begin{displaymath}
\left[F^{(BC)}_{S_2}(\mathbf{p}_2)\otimes F^{(CA)}_{S_1}(\mathbf{p}_1)\right]_{SM}
\left[F^{(bc)}_{S_2}(\mathbf{p}_2)\otimes F^{(ca)}_{S_1}(\mathbf{p}_1)\right]_{S-M}
\end{displaymath}
where $|S_1-S_2|\leq S \leq S_1+S_2$. We retain the
summation over the intermediate states as a suitable tool for the exploration of DCE form factors and 2BTD.
 {Considering, for instance, the $\{A\}$ system,} the two-body DCE form factors of Eq.(\ref{eq_PT}), once summed over the
intermediate states, are obtained as a superposition of
irreducible multipole components, related to DCE transitions of
orbital angular momentum $L$ and total spin $I_A$:
\bea
&&R^{J_AJ_BI_A}_{LS}(p_1,p_2) =\\
&&\sum_{\substack{I_1I_2J_C\\l_1l_2}} {\cal A}_{I_1I_2J_C}^{l_1l_2}
\mathcal{C}_{\ell_1,\ell_2}^L
{\cal R}^{J_BJ_C}_{\ell_2S_2I_2}(p_2){\cal R}^{J_CJ_A}_{\ell_1S_1I_1}(p_1)\nonumber.
\label{eq_PT_rid}
\eea
In the Supplemental Material \cite{SM} (Sections II and III), the recoupling techniques and the definitions of the reduced SCE matrix elements ${\cal R}^{J_DJ_E}_{\ell S I}(p)$ are discussed in due detail.

Changing coordinates $\{\mathbf{p}_1,\mathbf{p}_2\}\to \{\mathbf{q},\mathbf{v}\}$, where
$\mathbf{v} = 
\left(\mathbf{p}_1-\mathbf{p}_2\right)$ 
and $\mathbf{q} = \left(\mathbf{p}_1+\mathbf{p}_2\right)$, 
we perform the
$d^3v$ integral over a finite volume $V_v=\frac{4\pi\overline{v}^3}{3}$ of radius $\overline{v}$, thus introducing effectively a momentum cut-off. The resulting effective two-body transition form factor is
\begin{equation}\label{2BTD_xi}
F^{(BCA)}_{SM}(\boldsymbol{q}) \equiv \frac{(2\pi)^3}{V_v}\int d^3r\, e^{i\boldsymbol{q}\cdot\mathbf{r}}
[F_{S_2}^{(BC)}(\mathbf{r})\otimes F_{S_1}^{(CA)}(\mathbf{r})]_{S,M}
\end{equation}
which is determined by the multipole two-body transition densities (2BTD)
${\rho}^{J_AJ_BI_A}_{LS}(\mathbf{q})$:
\be
{\rho}^{J_AJ_BI_A}_{LS}(\mathbf{q})=
\frac{(2\pi)^3}{V_v}
\int d^3r\, e^{i\boldsymbol{q}\cdot\mathbf{r}}
[R^{J_AJ_BI_A}_{LS}(r,r) Y_{L M_{L}}(\hat{\mathbf{r}})].
\label{2BTD_red}
\ee
We will refer to Eq.(\ref{2BTD_xi}) and Eq.(\ref{2BTD_red}) as ``average approximation''  hereafter. The cut-off radius $\overline{v}$ is fixed by requiring that the nuclear matrix element $F^{(BCA)}_{SM}(\boldsymbol{q}=0)$ exhausts the transition strength by saturation 
(at $\approx$ 95$\%$),
reflecting the momentum structure of the nuclear 1BTD.
We find
$\overline{v} = 2.1$ fm$^{-1}$
and $\overline{v} = 2.2$ fm$^{-1}$ for projectile and target, respectively.

The whole DSCE transition matrix element (TME) can be recast to:
\begin{equation}
\label{T-matrix_like_SCE}
\begin{split}
&M^{(2)}_{\beta\alpha}(\mathbf{k}_\beta,\mathbf{k}_\alpha) \simeq
L^{(+)}_\gamma(\omega_\alpha)~
\sum_{S_1,S_2}\sum_{S,M}
(-1)^{S_1+S_2+S-M}\\
&\sum_{c,C}\int d^3q\, F^{(BCA)}_{SM}(\boldsymbol{q}) F^{(bca)}_{S-M}(\boldsymbol{q})
\tilde{V}_{S_1S_2}^{DSCE}({q})D_{\alpha\beta}(\boldsymbol{q}).
\end{split}
\end{equation}
The strengths of the components of the effective rank-2 iso-tensor interaction are determined by the vertex functionals:
\begin{equation}\label{V_NN_DSCE_AvrgRho}
\tilde{V}_{S_1S_2}^{DSCE}(q)\equiv
\int 4\pi r^2 dr\, j_0(qr) V_{S_2T}({r}) V_{S_1T}({r}).
\end{equation}
Further details on the angular momentum couplings
entering Eqs.(\ref{2BTD_xi}), (\ref{2BTD_red}) and (\ref{T-matrix_like_SCE})
are found in Section III of the Supplemental Material \cite{SM}.

It is worth noting that
Eq.\eqref{T-matrix_like_SCE} shows a TME structure very similar
to that of a single-step SCE transition, however with an effective four-body transition form factor
\begin{equation}
\begin{split}
&\mathcal{F}^{S_1,S_2}(\mathbf{q}) = \sum_{S,M}
(-1)^{S-M} \sum_CF^{(BCA)}_{SM}(\boldsymbol{q}) \sum_cF^{(bca)}_{S-M}(\boldsymbol{q})\\
\end{split}
\label{form_s}
\end{equation}
composed of the product of a pair of two-body DCE projectile and target form factors.
Uncertainties induced by
the cut-off radius $\bar{v}$ employed in "average approximation" can be kept under control by taking standard t-channel results for the global form factor at $\boldsymbol{q} \approx$ 0 as a reference.
It is noteworthy that the 2BTD of Eq. (\ref{2BTD_red}) are closely related to the nuclear form factors entering into DBD-NMEs as seen by comparing the formalism in
\cite{Barea-2015} and emphasized in recent attempts to connect spin-isospin nuclear excitation
strengths \cite{Roca2020,Ferreira2020,   Kauppinen:2022cii} to DBD-NMEs \cite{Shimizu2018,Santopinto2018,Ejiri:2022zdg,WangPLB2024}.

{\it Illustrative results.}
The essential features of the t-- and s--channel DSCE schemes are illustrated and compared for
the reaction $^{76}$Se ($^{18}$O, $^{18}$Ne$_{gs}$) $^{76}$Ge$_{gs}$ at $T_{lab}= 15$~MeV/A, currently investigated at LNS Catania \cite{NUMEN:2021ezg}.
Only $0^+\to 0^+$ transitions will be considered for both projectile and target. ISI and IFI, giving rise to
quenching effects in the DCE cross sections due to the absorption of the incoming probability flux \cite{Lenske:2024dsc},
are accounted for by properly chosen optical model potentials, according to the procedures described in \cite{Cappuzzello:2022ton}, see Section I of the Supplemental Material.
The projectile-target SCE interactions are treated by free-space NN isovector T-matrices in Love--Franey parametrization using the recently determined low--energy parameters, see e.g.
\cite{Lenske:2018jav,NUMEN:2021zst,Cappuzzello:2022ton}. Spin--scalar, spin--vector, and rank-2 spin-tensor interactions and microscopic transition form factors are used \cite{Lenske:2018jav}. Besides the genuine interest in higher order nuclear processes the chosen reaction is of relevance for DBD physics because $^{76}$Ge is a candidate for neutrinoless DBD.

Nuclear ground states and intermediate channel excited states
are described in Hartree--Fock plus BCS theory and Quasiparticle
Random Phase Approximation (QRPA). The self-consistent code developed by the Milano group \cite{Colo:2021ood} has been employed and  extended to treat CE excitations.   In the particle-hole channel, we consider three different Skyrme forces, namely SAMI \cite{SAMI}, SKX \cite{SKX} and SLy4 \cite{SLY4}, characterized by progressively increasing $G_0'$ Landau-Migdal parameter \cite{Towner}, i.e.,  0.35, 0.49 and 0.90, respectively. This parameter determines at leading order the spin-isospin properties of nuclear matter \cite{Bender},
thus allowing
to probe the sensitivity of our results to the choice of the interaction.
The Skyrme forces are supplemented by a density dependent contact interaction in the particle--particle channel, assuring proton and neutron pairing gaps  $\Delta_{p,n}\simeq 1$~MeV in $^{76}$Ge, and a neutron gap $\Delta_n\simeq 1$~MeV in $^{18}$O. 
Multipolarities up to $J^\pi=7^\pm$ are computed in configuration spaces large enough to guarantee that the  Ikeda--type sum rules \cite{Hara} are exhausted by better than 1$\%$. As widely practiced in DBD studies the TME of the second-step are approximated by QRPA SCE-TME
corresponding to transitions from the final nuclei  to the intermediate channel states. For the present purpose of a proof of principle study, we neglect the known non-orthogonality problems \cite{Pacearescu:2003ri}. An approach avoiding those problems was indicated  in \cite{Lenske:2021jnr}.

\begin{figure}
\includegraphics[scale=0.22]{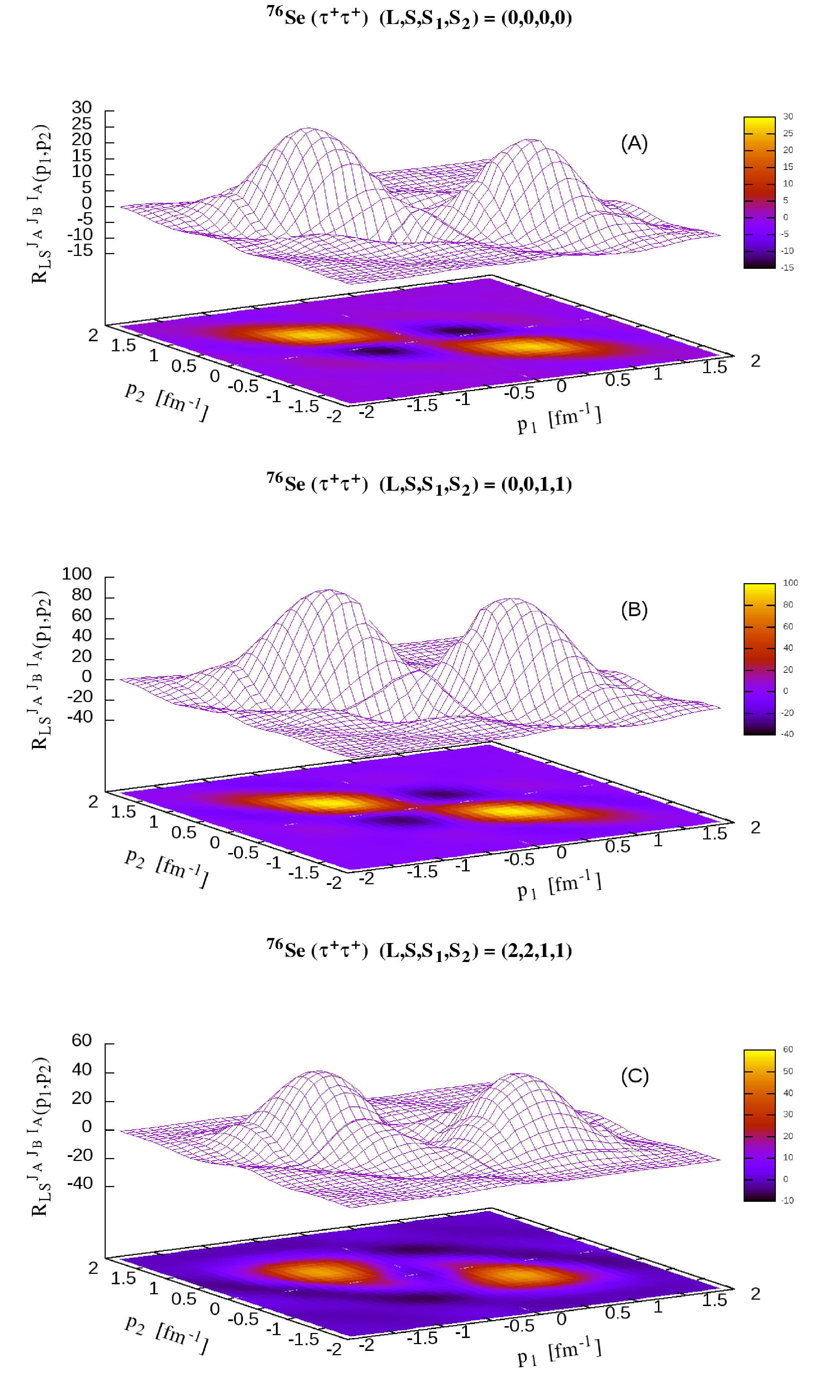}
\caption{2-body transition densities $R^{J_DJ_EI}_{LS}$ 
accounting for
target $^{76}$Se $\to ^{76}$Ge$_{gs}$ DCE transitions, as a function of the 
projections ($p_1$ and $p_2$) of the two linear momenta ${\bf p_1}$ and ${\bf p_2}$ along the total momentum transfer  ${\bf q}$. 
Each row illustrates the results for a given combination of $(L,S,S_1,S_2)$ angular momenta. }  \label{fig:2BTDs_vs_p1_p2}
\end{figure}

{\it Two-Body Transition Densities.} To get an overall picture of the nuclear DCE form factors of
Eqs (\ref{eq_PT},\ref{eq_PT_rid}),
Fig.\ref{fig:2BTDs_vs_p1_p2} represents, for the target case, 
the 
multipole components $R^{J_AJ_BI_A}_{LS}$
in terms of the projections of the momenta ${\bf p_1}$ and ${\bf p_2}$ along the total momentum transfer  ${\bf q}$.
The SLy4 force was employed in the calculations.
It is worth noting that relevant (positive or negative) contributions to this quantity
are located along the two bisectors. 
In particular, the behavior along the ${\bf v} = 0$ axis
corresponds to the
diagonal component of the 2BTD, with  ${\bf p_1} = {\bf p_2} = {\bf q}/2$.
On the other hand, the integral over ${\bf v}$
of $R^{J_AJ_BI_A}_{LS}(p_1,p_2)$ along the ${\bf q} = 0$ axis
(${\bf p_2} = -{\bf p_1}$)
is associated with the ${\bf q} = 0$ value of the 2BTD in ``average approximation", see Eq.(\ref{2BTD_red}).
Moreover, Fig. \ref{fig:2BTDs_vs_p1_p2} shows that significant contributions
to the two-body transition densities are essentially
confined to the region up to $|{\bf p_1}|=|{\bf p_2}|\approx 2$ fm$^{-1}$.
As expected, the main contributions at small linear momentum transfer values are due to Fermi-like and Gamow-Teller-like transitions (the upper and middle panels of Fig. \ref{fig:2BTDs_vs_p1_p2}), whereas contributions mostly associated with the rank-2 tensor components peak at higher momentum ($|{\bf p_1}|,|{\bf p_2}| \ge 0.5$ fm$^{-1}$)\footnote{According to our definitions, two-body transition
densities of the Gamow-Teller and Fermi operators
have the same sign.}.
However, because of the
compensation between positive and negative contributions when averaging over the momentum difference ${\bf v}$,
see for instance the positive and negative peaks appearing along the ${\bf q} = 0$ (${\bf p_2} = -{\bf p_1}$) axis,
2BTD associated with rank-2 tensor transitions turn out to be negligible.
In the calculations, we include intermediate states up to 50 MeV of excitation energy,
as the SCE transition stregths to the intermediate channel become negligible at higher energy.
\begin{figure}
\centering
\includegraphics[scale=0.22]{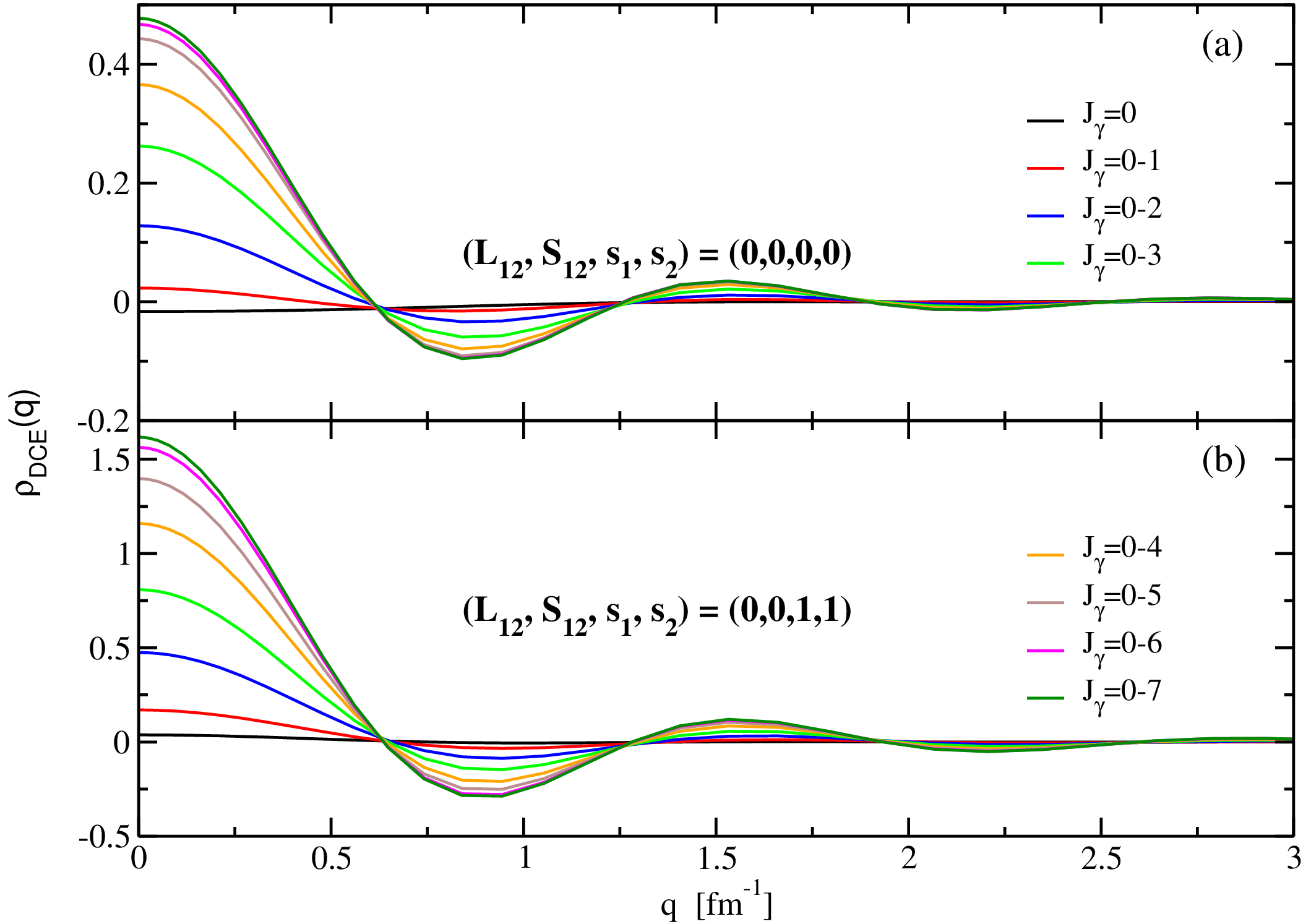}
\vskip 0.2cm
\includegraphics[scale=0.22]{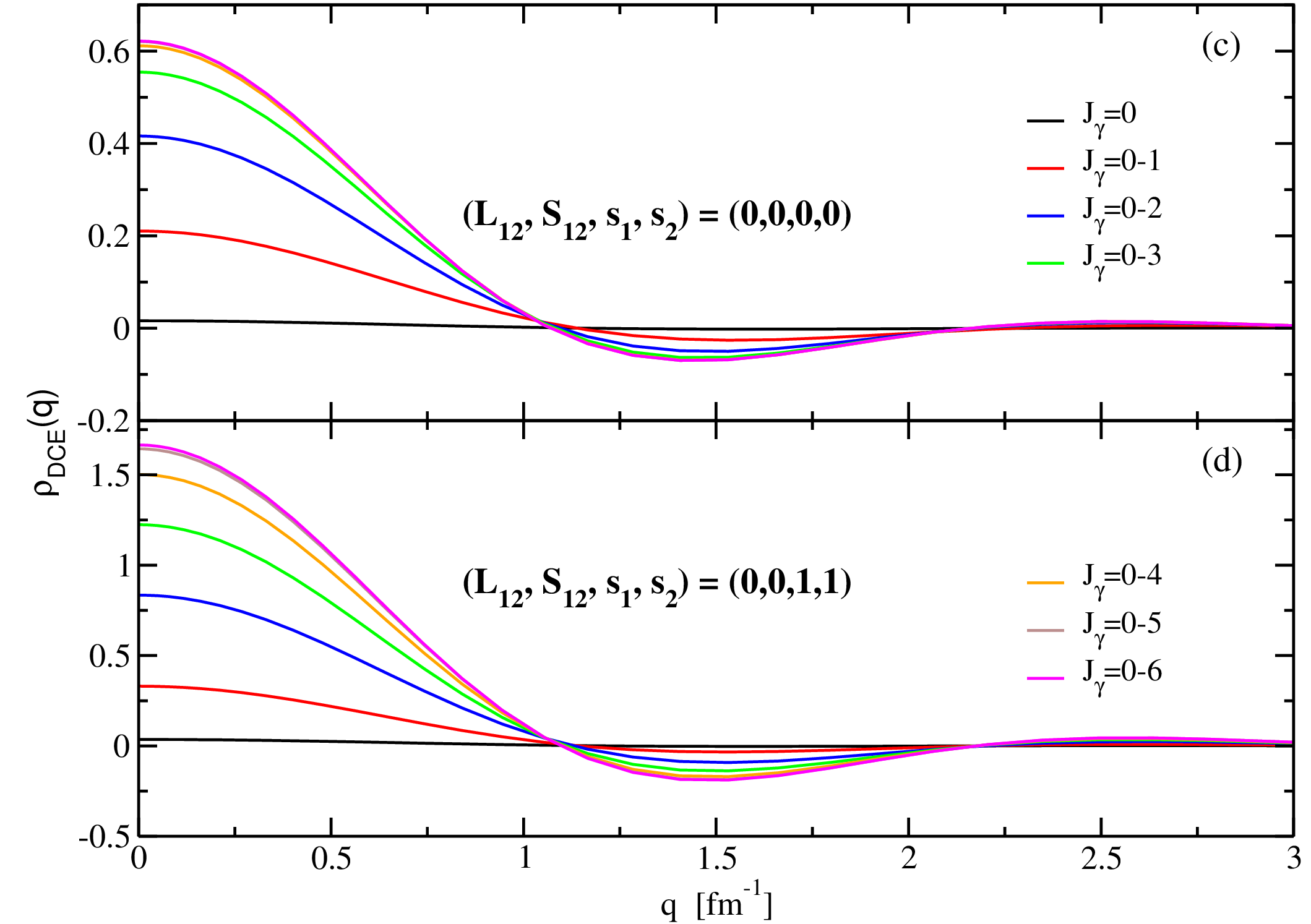}
\caption{2-body transition densities  ${\rho}^{J_DJ_EI}_{LS}(q)$, obtained in
"average approximation",
accounting for
projectile $^{18}$O $\to ^{18}$Ne$_{gs}$ (panels (c) and (d))
and target $^{76}$Se $\to ^{76}$Ge$_{gs}$   (panels (a) and (b)) DCE transitions, as a function of the two-step linear momentum transfer $q$. Each row illustrates the results for a given combination of $(L,S,S_1,S_2)$ angular momenta. The contributions from $(L,S,S_1,S_2)=(2,2,1,1)$ are omitted.}\label{fig:2BTD_avrgR_vs_q}
\end{figure}
Fig. \ref{fig:2BTD_avrgR_vs_q} illustrates {target} (panels (a) and (b)) and 
{projectile} (panels (c) and (d)) 2BTD, for different maximum values of the intermediate state spins $J_c,J_C$. In particular, we show only the Fermi-like $(L,S,S_1=0,S_2=0)$  and GT-like $(L,S,S_1=1,S_2=1)$ contributions, being the rank-2 tensor contributions negligible, where  $L$ and $S$ ($L=S=0$) are the total angular momentum and spin of the two-step transition feeding the final state ($J^\pi=0^+$).
As expected, relevant contributions to projectile and target 2BTD come from a considerably wide range of total angular momentum transfer.
Saturation is reached around $J_c=5$ and $J_C=6$ for projectile and target 2BTD, respectively, reflecting the size of the nucleus considered.

\begin{figure}
\centering
\includegraphics[width=8cm]{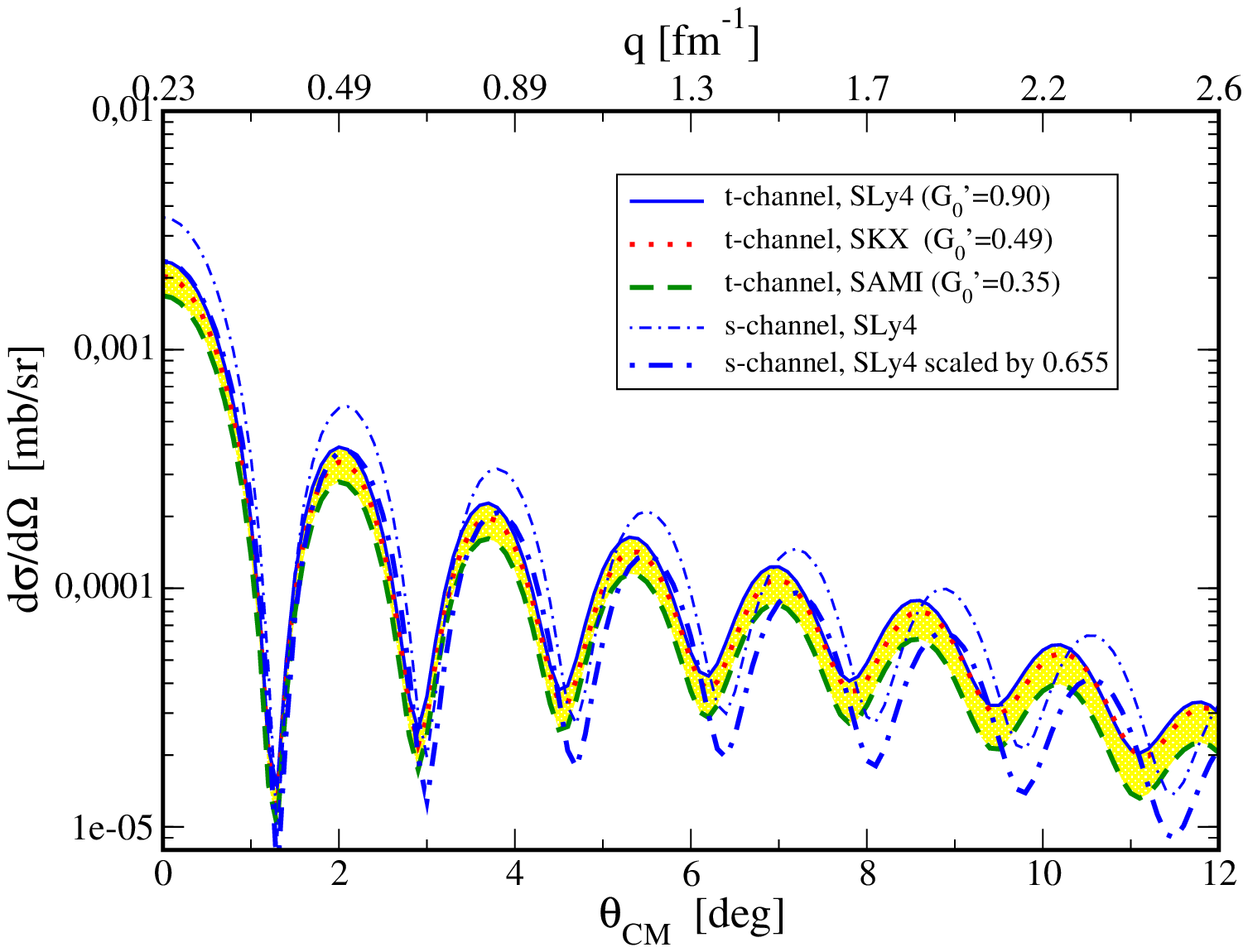}
\caption{Comparison between the DSCE angular distributions obtained in "average approximation" s-channel and 
in t-channel representations, with the Sly4 interaction. The yellow band indicates the spread of t-channel results adopting different Skyrme interactions. On the upper abscissa the momentum transfer is displayed. 
}\label{fig:FF_mod_t-_vs_s-channel}
\end{figure}

{\it DSCE cross section: s-channel vs t-channel.} 
The differential cross section 
obtained in s-channel "average approximation" with the SLy4 parametrization
is shown
in Fig.\ref{fig:FF_mod_t-_vs_s-channel},
together with corresponding t-channel results.
In {all} calculations, an overall
normalization, corresponding to a channel-independent $|L^{(+)}_\gamma(\omega_\alpha)| \approx 1/300~MeV^{-1}$ was
employed, to reproduce, within the t-channel scheme, full DW results, as in  Ref. \cite{Bellone:2020lal}.  

The s-channel results overestimate, by about 35$\%$, t-channel calculations at forward angles (similar discrepancies are observed using other Skyrme interactions).
However, scaling the s-channel curve to the t-channel results at $\theta$ = 0,
a satisfactory agreement is seen between the two calculations up to $\theta \sim 5^\circ$, namely for momentum transfers
$k_{\alpha\beta}< 200$~MeV/c ($k_{\alpha\beta}< 1fm^{-1}$). The difference between the two curves keeps smaller than (or at most comparable to) the sensitivity of the DCE cross section 
to the Skyrme interactions considered in our study, represented by the yellow band in the figure. It is interesting to note that larger $G'_0$ values lead to increasing DCE cross sections.
At larger angles, the oscillatory patterns of s-channel and t-channel calculations run increasingly out of phase.

These discrepancies could be attributed to the fact that
the s-channel ``average'' procedure implies a larger contribution of high multipolarity transitions within the off-shell intermediate channel, especially
associated with the heavier system. Based on the factorization introduced in s-channel, more accurate results are expected for reaction systems with greater mass asymmetry,  as, in that case, the form factors in both the t-channel--combining projectile and target at each step--and the separation approach of the s-channel 
are predominantly governed by the size of the heavier reactant.

{\it Conclusions.} The spectroscopy of higher order nuclear processes is a highly interesting although rarely studied
area of research. One reason is the lack of appropriate probes with the proper operator structure to excite in
laboratory experiments selectively higher order nuclear modes  of interest. In this Letter we have pointed out the
large, albeit to be further explored, research potential of heavy ion DCE reactions for second order nuclear
spectroscopy. A key role is played by collisional DCE reaction for which DSCE reactions are a representative example. A
new DSCE formulation was presented with the central result of the clear formal separation of  probe-independent
spectroscopic 2BTD structures from the probe-dependent parts, as ISI/FSI and the interaction vertex functionals. In the proposed formulation, the
latter appear as renormalization factors of the bare nuclear polarization tensors.
Finally, it is worth mentioning that in a physical reaction the DSCE process might compete
with direct DCE mechanisms relying on virtual ($\pi^\pm,\pi^\mp$) reactions, namely the Majorana DCE (MDCE) mechanism described in \cite{Lenske:2024mdc}. Although DSCE and MDCE processes proceed by vastly different operators, as far as the
spectroscopy is concerned, the two distinct reaction mechanisms are determined by the same 2BTD discussed
here.

\section{Acknowledgments}
We thank C. Agodi, F. Cappuzzello and M. Cavallaro for fruitful discussions. D.G. thanks G. Col\'o  and X. Roca-Maza for useful discussions and support in the use of Skyrme-QRPA code.
{Partial funding of the NUMEN collaboration by the European Union's Horizon 2020 research and innovation program under Grant No. 714625 is acknowledged. H.L. acknowledges gratefully partial support by Alexander-von-Humboldt Foundation, DFG, grant Le439/17, and INFN.}

\newpage

\end{document}